\documentclass[a4paper,11pt]{article}
\usepackage{jheppub} 
\usepackage{etoolbox}
\patchcmd{\maketitle}{\@fpheader}{}{}{}
\usepackage[T1]{fontenc} 
\usepackage[utf8]{inputenc} 
\usepackage{microtype} 
\usepackage{lmodern} 
\usepackage{mdframed} 
\usepackage{mathtools}
\usepackage{graphicx}
\usepackage{cancel} 
\usepackage{todonotes} 
\usepackage[normalem]{ulem} 
\usepackage{soul} 
\usepackage{amsmath}
\usepackage{amsfonts}
\usepackage{amssymb}
\usepackage{stmaryrd}
\usepackage[mathscr]{euscript}
\usepackage{mathtools}
\usepackage{mathrsfs}
\usepackage{multicol}
\usepackage{mleftright} \mleftright
\usepackage{tensor} 
\usepackage{braket} 
\usepackage{mathrsfs} 
\usepackage{bbold}
\usepackage{color}
\usepackage{braket}
\usepackage{slashed}
\usepackage{hyperref}

\newcommand{\comment}[1]{}

\providecommand*{\dd}{\mathop{}\!\mathrm{d}}
\renewcommand*{\dd}{\mathop{}\!\mathrm{d}}

\providecommand{\sc}{{ }}
\renewcommand{\sc}{{ }}

\DeclareMathAlphabet{\mathfs}{U}{rsfs}{m}{n}                     %

\newcommand{\inter}{{\lrcorner}}

\newcommand{\be}{\nopagebreak[3]\begin{equation}}
\newcommand{\ee}{\end{equation}}
\newcommand{\bee}{\nopagebreak[3]\begin{equation*}}
\newcommand{\eee}{\end{equation*}}
\newcommand{\ba}{\nopagebreak[3]\begin{eqnarray}}
\newcommand{\ea}{\end{eqnarray}}
\newcommand{\baa}{\nopagebreak[3]\begin{eqnarray*}}
\newcommand{\eaa}{\end{eqnarray*}}
\newcommand{\bal}{\nopagebreak[3]\begin{aligned}}
\newcommand{\eal}{\end{aligned}}

\newcommand{\bseq}{\nopagebreak[3]\begin{subequations}}
\newcommand{\eseq}{\end{subequations}\noindent}

\title{Thermodynamics of the three-dimensional black hole with torsion}
\author[1]{Luis Avilés}
\author[2]{Diego Hidalgo}
\author[1]{Omar Valdivia}

\affiliation[1]{Instituto de Ciencias Exactas y Naturales (ICEN), Universidad Arturo Prat,\\ Avenida Arturo Prat Chacón 2120, 1110939, Iquique, Chile}
\affiliation[2]{Science Institute, University of Iceland,\\ Dunhaga 3, 107 Reykjav\'ik, Iceland}

\emailAdd{luaviles@unap.cl}
\emailAdd{dhidalgo@hi.is}
\emailAdd{ovaldivi@unap.cl}
\preprint{{\bf } }

\abstract{The stationary black hole solution of a Chern-Simons model based on the semi-simple extension of the Poincaré gauge group is studied. The solution resembles the metric properties of the BTZ geometry but contains, in addition, non-vanishing torsion. The global structure of spacetime is characterized by three conserved charges: two associated with the mass and angular momentum and one extra constant triggered by spacetime torsion. Consequently, we show that the entropy deviates from the standard Bekenstein-Hawking value and discuss the implications of torsional charges in the context of black hole thermodynamics.}

\begin{document}
\maketitle
\flushbottom

\section{Introduction}\label{intro}
Thermodynamic properties of black holes are expected to give us important clues in understanding the quantum nature of gravity~\cite{PhysRevD.7.2333,Hawking:1972aa,Bardeen:1973aa}. The development of black hole thermodynamics has involved many significant results, such as the intimate connection among the laws of gravitation with the laws of thermodynamics \cite{Jacobson:1995ab,Padmanabhan:2009vy}, intrinsic relationships between black hole entropy and Noether charges associated with diffeomorphism symmetry \cite{Wald:1993nt}, and thermal phase transitions due to the presence of a negative cosmological constant \cite{Hawking:1983aa}. 
In this regard, the discovery of the three-dimensional black hole by Bañados, Teitelboim, and Zanelli (BTZ)~\cite{Banados:1992wn,Banados:1992gq} was of particular importance since it allowed us to investigate in a more straightforward way the nature of black holes and their relationship to basic foundations of quantum physics. Further developments of three-dimensional black holes can be found in~\cite{Carlip:1995qv,Martinez:1996gn,Carlip:1998wz,Cardenas:2014kaa,Perez:2015kea,Erices:2017nta} and references therein.

It is well known that three-dimensional gravity with a negative cosmological constant can be formulated as a Chern-Simons (CS) theory for the anti-de Sitter group $SO(2,2)$~\cite{ACHUCARRO198689,WITTEN198846}. The gauge field takes values in the  Lie algebra $\mathfrak{so}(2,2)$ and is written in terms of the dreibein $e^{a}(x)$ and the spin connection $\omega^{a}(x)$, as independent degrees of freedom. The equations of motion imply constant (Riemannian) curvature and vanishing torsion, so the theory is purely metric on-shell. However, there are interesting trends motivated by gauge-theoretic arguments for considering gravity models based on Riemann-Cartan geometry
containing both metric and affine properties of spacetime as generic ingredients of the
gravitational dynamics (see, for instance,~\cite{RevModPhys.48.393,Hehl:1994ue,Blagojevic:2013xpa}, and references therein). 
 In three-dimensions, Riemann-Cartan gravity is studied in \cite{MIELKE1991399}. Further development along these lines led to several interesting results, such as a stationary black hole solution \cite{Garcia:2003nm}, asymptotic symmetries, thermodynamic properties, and microscopic description of the entropy \cite{Blagojevic:2006jk,Blagojevic:2003uc,Blagojevic:2006hh}. 
  An important observation in this context is that entropy differs from the Bekenstein-Hawking result by an additional contribution, which depends on the presence of spacetime torsion.
 
 In this work, we study the thermodynamic properties of another black hole solution with torsion, originally reported in~\cite{Hoseinzadeh:2014bla}. The theory behind this solution is a three-dimensional CS model genuinely invariant under the semi-simple extension of the Poincaré gauge symmetry~\cite{Soroka:2006aj}.  
 This semi-simple enlargement is equivalent to an S-expansion procedure of the AdS algebra in any dimension and is often referred to as AdS-Lorentz (AdSL) Lie algebra~\cite{Izaurieta:2006zz,Diaz:2012zza,Fierro:2014lka}. The AdSL group was introduced to describe the symmetries of a particle moving in AdS spacetime in the presence of a constant electromagnetic field \cite{Gomis:2009dm}. Moreover, AdSL symmetry is also related to the so-called Maxwell group~\cite{Schrader:1972zd} through an Inönü-Wigner contraction. In the context of CS theories of gravity, expanded symmetry algebras have many interesting properties. For instance, any expanded algebra may include new generators, increasing the gauge symmetry of the particular theory. In such cases, it also means including compensating gauge fields apart from the dreibein and spin connection. Interestingly, in three-dimensions, this extra gauge field codifies the affine structure of geometry so that gauge invariance is preserved.\footnote{There are other interesting examples of black hole solutions for theories with extended symmetries, such as higher spin systems~\cite{Blencowe:1988gj,Bergshoeff:1989ns,Vasiliev:1995dn,Perez:2012cf}.  }~\cite{Diaz:2012zza,Salgado:2014qqa}.
 
Here, we start by highlighting the fact that three-dimensional AdSL gravity naturally contains torsion. Moreover, we show that the stationary solution~\cite{Hoseinzadeh:2014bla} is characterized by the standard constants of motion $M, J,$ of Einstein's gravity, 
and a new constant $b$ related to the presence of spacetime torsion. The presence of this constant generalizes the \textquotedblleft{}exotic'' thermodynamics analysis of Townsend and Zhang~\cite{Townsend:2013ela} by including the torsion in both the global charge and entropy formulas. 
Specifically, we present the modifications in entropy caused by torsion and the additional term in the first law, which comes from the new global charge. For completeness, we display the Smarr formula with the effects of torsion taken into consideration. For this purpose, we employ the methods used in the CS formalism proposed in \cite{Carlip:1994gc,Henneaux:2013dra,Bunster:2014mua}. 

This paper is organized as follows. In Section \ref{Section:Symmetry}, we review the AdSL symmetry leading properties and the construction of a CS principle action. In Section \ref{Section:StationanryBH}, we explain the stationary solution of this theory and review the BMS-like gauge for the field content that allows solving the field equations. In Section \ref{Section:asymptoticsymmetry}, we review the corresponding asymptotic symmetry and the derivation of the asymptotic gauge connection useful for the computation of the entropy. The study of the computation of the global charges, entropy, first law, and Smarr formula is developed in Section \ref{Section:Thermo}. Finally, Section \ref{Section:conclusions} is devoted to discussing the results and future directions. 

\section{Three-dimensional AdS-Lorentz gravity}
This section briefly reviews the formulation of a CS theory of gravity quasi invariant under the AdSL group \cite{Diaz:2012zza}. Some subtleties in deriving the black hole solution of this gravity theory are highlighted. 
\subsection{Formulation of AdS-Lorentz CS gravity} \label{Section:Symmetry}
In $(2+1)$-spacetime dimensions, the AdSL algebra is generated by spacetime rotations $J_{a}$, spacetime translations $P_{a}$, and a new type of generators $Z_{a}$, originally introduced in \cite{Schrader:1972zd,Bacry:1970aa}. Commutation relations among these generators are given by
\begin{equation}
\begin{tabular}{lll}
 \ensuremath{\left[J_{a},J_{b}\right]=\epsilon_{abc}J^{c}\,,}  &  \medskip &  \ensuremath{\left[Z_{a},Z_{b}\right]=\epsilon_{abc}Z^{c}\,},\\
 \ensuremath{\left[J_{a},P_{b}\right]=\epsilon_{abc}P^{c}\,,}  & \medskip   &  \ensuremath{\left[J_{a},Z_{b}\right]=\epsilon_{abc}Z^{c}}\,,\\
 \ensuremath{\left[P_{a},P_{b}\right]=\epsilon_{abc}Z^{c}}\,,  &  \medskip  &  \ensuremath{\left[Z_{a},P_{b}\right]=\epsilon_{abc}P^{c}}\,,
\end{tabular}\label{adsl} 
\end{equation}
where $\epsilon_{abc}$ is the Levi-Civita symbol, such that $\epsilon_{012}=1, \, \epsilon^{012}=-1$, and $a=0,1,2$ is a Lorentz index, raised and lowered with the Minkowski metric $\eta_{ ab} = \text{diag} (-1,+1,+1)$. As presented in (\ref{adsl}), is clear that generators $Z^{a}$ complete the semi-simple extension of the Poincaré algebra $\mathfrak{iso}(1,2)$ generated by rotations $J_{a}$ and translations $P_{a}$.
Another observation is that \eqref{adsl} can also be written as the direct sum $\mathfrak{so}(2,2)\oplus\mathfrak{so}(2,1)$, reason why it is usually referred to as AdSL algebra. In fact, defining
\begin{equation}\label{tilde}
	\tilde{J}_{a}  =  Z_{a}\,, \hspace{1cm}\tilde{P}_{a} = P_{a}\,, \hspace{1cm} \tilde{Z}_{a}  =  J_{a}-Z_{a}\,,
\end{equation}
the direct sum structure of AdS and Lorentz algebras is made explicit (see \cite{Diaz:2012zza,Salgado:2014qqa}). This last observation also implies that (\ref{adsl}) is equivalent to the direct sum of three copies of the Lorentz algebra. Indeed, three commuting sets of $\mathfrak{so}(2,1)$ generators
\begin{equation} 
\left[J_{a}^{+}, \, J_{b}^{+}\right]= \epsilon_{abc} (J^{+})^{c}\,,\hspace{1.2cm} \left[J_{a}^{-}, \, J_{b}^{-}\right]= \epsilon_{abc}(J^{-})^{ c}\,, \hspace{1.2cm} 
\left[\hat{J}_{a}, \, \hat{J}_{b}\right] = \epsilon_{abc}\hat{J}^{c}\,,
\end{equation}
reproduce \eqref{adsl} by means of redefinitions
\begin{equation}
Z_{a}  =  J_{a}^{+}+J_{a}^{-} \,,\hspace{1.2cm}
P_{a}  =  J_{a}^{+}-J_{a}^{-} \,, \hspace{1.2cm}
J_{a}  =  \hat{J}_{a}+J_{a}^{+}+J_{a}^{-}\,.
\end{equation}
The most general  invariant tensor associated to \eqref{adsl} has the following components
\begin{subequations}\label{Eq:AdSLorentzpairing}
\begin{eqnarray}
\left\langle J_{a}\,,J_{b}\right\rangle =\alpha_{0}\eta_{ab}\,, & \bigskip  \hspace{1cm}& \left\langle P_{a}\,,P_{b}\right\rangle = \alpha_{2}\eta_{ab}\,, \\
\left\langle J_{a}\,,P_{b}\right\rangle = \alpha_{1}\eta_{ab}\,, & \bigskip\hspace{1cm}  & \left\langle Z_{a}\,,Z_{b}\right\rangle = \alpha_2 \eta_{ab}\,,\label{eq:invt}\\
\left\langle J_{a}\,,Z_{b}\right\rangle = \alpha_{2} \eta_{ab}\,, & \bigskip  \hspace{1cm}& \left\langle Z_{a}\,,P_{b}\right\rangle = \alpha_1 \eta_{ab}\,,
\end{eqnarray}
\end{subequations}
where $\alpha_{0},$ $\,\alpha_{1}, \text{ and }\alpha_{2}$ are real arbitrary constants. This bilinear form is non-degenerate if $\alpha_{1}\neq \alpha_{2}$ and $\alpha_{2} \neq \alpha_{0}$.   The Cartan sub-algebra is a three-dimensional space which suggest the presence of three independent Casimir operators constructed from these generators \cite{Diaz:2012zza,Fierro:2014lka}.

A quasi-invariant gravity theory with AdSL symmetry can be constructed by evaluating the CS action
\begin{equation}
	I\left[A\right]=\frac{k}{4\pi}\int_{\scriptscriptstyle \mathcal{M}}\left\langle A\wedge \mathrm{d}A+\frac{2}{3}\,A\wedge A \wedge A \right\rangle\,,\label{gcs}
\end{equation}
for the connection one-form $A=A_{\mu} \dd x^{\mu}$, taking values in the algebra \eqref{adsl} spanned by $\{P_a,J_a,Z_a \}$
\begin{equation}
A=\frac{1}{\ell} e^{a}P_{a}+\omega^{a}J_{a}+\sigma^{a}Z_{a}\,.\label{1f}
\end{equation}
As usual, $e^{a}(x)$ denote the one-form dreibein, $\omega^{a}(x)$ is the one-form spin connection, and the one-form $\sigma^{a}(x)$ is the compensating gauge field associated with the non-abelian generator $Z_{a}$. Here, $\ell$ denotes the AdS radius and $k$ is the CS level related to the Newton's constant $G$ according to  $k=\frac{\ell}{4G_{N}}$. 
In terms of the gauge field components, the action takes the form\footnote{Henceforth, the wedge product $\wedge$ between differential forms is 
understood, i.e. $\omega^{a} \, e^{b} = \omega^{a}\wedge e^{b} = -e^{b}\wedge \omega^{a}.$}
\begin{equation}\label{cs}
I_{\scriptscriptstyle \mathrm{AdSL}}[e,\omega,\sigma]  = \frac{k}{4\pi}\int _{\scriptscriptstyle \mathcal{M}} \left(L_{\text{EC}} + L_{\text{exotic}}  + L_{\sigma} \right)\,, 
\end{equation}
with
\begin{eqnarray}
L_{\mathrm{EC}} & =&  \frac{\alpha_{1}}{\ell} \left( 2 R_{a}e^{a}+ \frac{1}{3\ell^{2}} \epsilon^{abc} e_a e_b e_c\right)\,,\\
L_{\mathrm{exotic}}   &=& \alpha_{0} \left( \omega^{a}\dd \omega_{a}+\frac{1}{3}\epsilon^{abc}\omega_{a}\omega_{b}\omega_{c} \right)+\frac{\alpha_{2}}{\ell^{2}}T^{a}e_{a}\,,\\
L_{\sigma}  & =& \frac{2\alpha_{1}}{\ell}F^{a}e_{a}+\alpha_{2}\left( 2R^{a}\sigma_{a}+\frac{1}{3}\epsilon^{abc}\sigma_a\sigma_b\sigma_c+ \mathrm{D}_{\omega}\sigma^{a}\sigma_{a}+\frac{1}{\ell^{2}}\epsilon^{abc}e_a\sigma_{b}e_{c}\right)\,.
\end{eqnarray}
The two-form curvatures associated to the gauge fields $\omega^a$, $e^a$ and $\sigma^a$ are defined by
\begin{equation}
R^{a}=\dd\omega^{a}-\frac{1}{2}\epsilon^{abc}\omega_{b}\omega_{c}\,,\hspace{1cm} T^{a}= \mathrm{D}_{\omega}e^{a}\,, \hspace{1cm} F^{a}= \mathrm{D}_{\omega}\sigma^{a}+\frac{1}{2}\epsilon^{abc}\sigma_{b}\sigma_{c}\,,
\end{equation}
where the covariant derivative is defined with respect to the full spin connection as $\mathrm{D}_{\omega}e^{a}\equiv \mathrm{d} e^{a}-\epsilon^{abc}\omega_{b}e_{c}$. The total Lagrangian in (\ref{cs}) consists of three pieces: the three-form $L_{\text{EC}}$ is the first order generalization of the Einstein-Hilbert term with a negative cosmological constant. The piece $L_{\text{exotic}} $ contains the exotic Lorentz-Chern-Simons term~\cite{DESER1982372} and a torsional contribution whose exterior derivative is locally related to the Nieh-Yan topological invariant~\cite{Nieh:1981ww,Chandia:1997jf}. The Lagrangian three-form $L_{\sigma}$ is the extra contribution coming from the presence of the gauge field $\sigma^{a}$. 

Functional variation of \eqref{cs} leads to the field equations
  \begin{eqnarray}
\delta\omega^{a} & : &   \qquad0=\alpha_{0}R_{a}+\frac{ \alpha_{1}}{\ell}\left(T_{a}+ \epsilon_{abc}\sigma^{b}e^{c}\right)+ \alpha_{2}\left(F_{a}+\frac{1}{2\ell^2}\,\epsilon_{abc}e^{b}e^{c}\right)\,,\nonumber \\ \medskip
\delta e^{a} & : &   \qquad0=\frac{\alpha_{1}}{\ell}\left(R_{a}+ F_{a}+\frac{1}{2\ell^{2}}\,\epsilon_{abc}e^{b}e^{c}\right)+\frac{\alpha_{2}}{\ell^2}\left(T_{a}+ \epsilon_{abc}\sigma^{b}e^{c}\right)\,,\label{eom}\\ \medskip
\delta\sigma^{a} & : &   \qquad0= \frac{\alpha_{1}}{\ell}\,\left(T_{a}+\epsilon_{abc}\sigma^{b}e^{c}\right)+\alpha_{2}\left(R_{a}+ F_{a}+\frac{1}{2\ell^{2}}\,\epsilon_{abc}e^{b}e^{c}\right)\,,\nonumber
\end{eqnarray}
so in principle the space of solutions contains different branches controlled by the possible values of the coupling constants $\alpha_0$, $\alpha_1$ and $\alpha_2$. However, requiring a non-vanishing determinant of (\ref{eom}), the equations of motion reduce to
\begin{subequations}\label{eomf}
\begin{eqnarray}
R_{a} & = & 0\,,\\
T_{a}+\epsilon_{abc}\sigma^{b}e^{c} & = & 0\,,\label{Eq:Contorsionequation}\\
F_{a}+\frac{1}{2 \ell^2}\,\epsilon_{abc}e^{b}e^{c} & = & 0\label{Eqct}\,.
\end{eqnarray}
\end{subequations}
This is, in fact, the generic case where the field equations completely determine the evolution of the system. Notice that the full dynamics of the theory is captured by the $\alpha_{2}$ term in the action \eqref{cs}, regardless of the choices for $\alpha_{0}$ and $\alpha_{1}$. Otherwise, $\omega^{a}$ would be a redundant gauge field of the theory, and therefore the bilinear form \eqref{Eq:AdSLorentzpairing} would be degenerate. Another important observation is that, in this particular dimension, the gauge field $\sigma^{a}$ is a source for torsion. This can be seen explicitly in Eq.(\ref{Eq:Contorsionequation}) by splitting the Lorentz connection $\omega^{a}$ in terms of its torsion-free $(\mathring{\omega}^{a})$ and contorsion part $(\kappa^{a})$ as $\omega^{a} = \mathring{\omega}^{a} + \kappa^{a}$ and then one intermediately arrives to $\sigma^{a}=-\kappa^{a}$. This last result makes the theory particularly appealing, for instance, when studying black hole solutions and their thermodynamics, as we will see later.
\subsection{Stationary black hole solution} \label{Section:StationanryBH}
Three-dimensional AdSL gravity admits a stationary black hole solution with rotation \cite{Hoseinzadeh:2014bla}. The line element is given by\footnote{Henceforth, we will use the convention $8G_{\scriptscriptstyle N}=1$ with $G_{\scriptscriptstyle N}$ being the Newton's constant.}
\begin{equation}
\dd s^{2}=-N^{2}\dd t^{2}+\frac{dr^{2}}{N^{2}}+r^{2}\left( \dd \varphi +N_{\varphi
}\dd t\right) ^{2}\,,  \label{ADM}
\end{equation}%
with the $r$-dependent functions
\begin{equation}
N^{2}=-M+\frac{J^{2}}{4r^{2}}+\frac{r^{2}}{\ell ^{2}}\,,\qquad N_{\varphi }=-
\frac{J}{2r^{2}}\,,
\label{N_and_Nphi}
\end{equation}
and $M$ and $J$ are integration constants. This means the BTZ tetrad\footnote{From now on, we will use bar notation both for the diagonal Minkowski metric $\overline{\eta}=\text{diag}(-1,1,1)$ and gauge fields associated.}
\begin{align}
	\bar{e}^{0}& =N\dd t\,,  \notag \\
	\bar{e}^{1}& =N^{-1}\dd r\,,  \label{btz1} \\
	\bar{e}^{2}& =r\left( \dd\varphi +N_{\varphi }\dd t\right) \,,  \notag
\end{align}
is a solution while the torsionless part of the spin connection $\mathring{\bar\omega}^{a}$ is solved from $\mathrm{D}_{\mathring{\omega}} {e}^{a}=0$
	\begin{align}
	\mathring{\bar\omega}^{0}  & =N\dd\varphi\,,\nonumber\\ 
	\mathring{\bar\omega}^{1}  & =-\frac{N_{\varphi}}{N}\dd r\,,\label{btz2}\\
	\mathring{\bar\omega}^{2}  & =r\left(  \frac{\dd t}{\ell^2}+N_{\varphi}\dd\varphi\right)\,.
	\nonumber
\end{align}
Expressions (\ref{btz1}) and (\ref{btz2}) determine completely the metric part of the solution. On the other hand, inserting $\sigma^{a}=-\kappa^{a}$ and (\ref{btz1})-(\ref{btz2}) into (\ref{Eqct}) one finds the following components for the contorsion tensor
\begin{align}
	\bar{\kappa}^{0}& =  a C \dd t+B \dd r+(C -N)\dd\varphi 
	\,,  \notag  \medskip \\
	\bar{\kappa}^{1}& = a F \dd t+\left( E +\frac{N_{\varphi }}{N}%
	\right) \dd r+ F \dd\varphi \,\,,  \label{btz3} \\
	\bar{\kappa}^{2}& =  \left(a I -\frac{r}{\ell ^{2}} \right)
	\dd t+H \dd r+\left(I-r N_{\varphi }\right)\dd\varphi  \,,  \notag
\end{align}%
where we have defined the $r$-dependent functions
\begin{eqnarray}
	B &\equiv &\frac{1}{I}\frac{\dd F}{\dd r}+\frac{H C}{I}\,, \medskip \notag \\
	C &\equiv &\sqrt{F^{2}+I^{2}+b\,},\medskip \label{BCE} \\
	E &\equiv &\frac{ 1}{C}\left(\frac{\dd I}{\dd r}+\frac{F}{I }\frac{\dd F}{\dd r}\right)+\frac{F H}{I}\,.  \notag
\end{eqnarray}
Therefore, the solution depends on three arbitrary functions $I=I(r)$, $F=F(r)$ and $H=H(r)$ and two additional integration constants $a$ and $b$.

Let us write down this solution in the BMS gauge. To this end, we introduce the off-diagonal Minkowski metric
\begin{equation}
	\eta_{ab}=\left(\begin{array}{ccc}
	0 & 1 & 0\\
	1 & 0 & 0\\
	0 & 0 & 1
	\end{array}\right)\,, \label{eta}
\end{equation}
and solve field equations in the direct sum basis \eqref{tilde}. By doing so, the torsionless fields $\tilde{e}^{a}$ and $\tilde{\omega}^{a}$ can be set to obey standard pure gravity boundary conditions, and the gauge field $\tilde{\sigma}^{a}$ is simply a flat Lorentz connection~\cite{Concha:2018jjj}. In the BMS gauge \cite{Barnich:2013yka,Barnich:2012aw}, the solution is parametrized by the local coordinates $x^\mu=(u,r,\phi)$, where $u=t-f(r)$ and $\phi = \varphi+ g(r)$ correspond to the retarded time coordinate and new angular coordinate, respectively, with
\begin{equation}
\frac{\dd f(r)}{\dd r} = N^{-2}\,,~~~~\frac{\dd g(r)}{\dd r} = - \frac{N_{\varphi}}{N^{2}}\,.
\end{equation}
The boundary is located at $r=const.$ With this new basis and change of variables, the metric \eqref{ADM} reads
\begin{equation}\label{BMS3}
\dd s^{2}=\eta_{ab}\tilde{e}^{a}\tilde{e}^{b}=2\tilde{e}^{0}\tilde{e}^{1}+\left( \tilde{e}^{2}\right) ^{2}\,,
\end{equation}%
and the dreibein can be chosen as 
\begin{subequations}
\begin{eqnarray}
\tilde{e}^{0} & =& \frac{1}{2\ell^{2}}\left(\ell^{2}\,\mathcal{M}(u,\phi) -r^{2}\right) \dd u- \dd r+ \frac{1}{2}\mathcal{N}(u,\phi) \dd \phi \,, \\
\tilde{e}^{1} & =&\dd u\,, \\
\tilde{e}^{2} & =&r\, \dd\phi \,.
\end{eqnarray}
\end{subequations}
Here, $\mathcal{M}$ and $\mathcal{N}$ are arbitrary functions to be fixed by the Einstein field equations. By considering the AdSL algebra as the direct sum $\mathfrak{so}(2,2)\oplus \mathfrak{so}(2,1)$, the spin connection $\tilde{\omega}^a$ is torsionless and therefore is given by
\begin{subequations}
\begin{eqnarray}
\tilde{\omega}^{0} &=& \dfrac{1}{2\ell ^{2}}\mathcal{N}(u,\phi) \dd u+ \frac{1}{2\ell^{2}}\left(\ell^{2}\, \mathcal{M}(u,\phi) -r^{2} \right) \dd\phi \,, \\
\tilde{\omega}^{1} & = &\dd\phi \,, \\
\tilde{\omega}^{2} & =&\frac{r}{\ell ^{2}}\dd u\,.
\end{eqnarray}
\end{subequations}
Finally, the gauge field $\tilde{\sigma}^{a}$ acquires the form
\begin{subequations}
\begin{eqnarray}
\tilde{\sigma}^{0}& =&\dfrac{1}{2\ell^{2}}\left(\ell^{2}\, \mathcal{M}(u,\phi)-\mathcal{R}(u,\phi)\right)\dd\phi \,,  \\
\tilde{\sigma}^{1}& =&\dd\phi \,,  \label{BMS6} \\
\tilde{\sigma}^{2}& =&0\,.  
\end{eqnarray}
\end{subequations}
with $\mathcal{R}$ another arbitrary function. 
By inserting this BMS gauge solution into the field equations \eqref{eomf}, we get (see, for instance, \cite{Barnich:2013yka})
\begin{equation}
\mathcal{\dot{M}}(u,\phi ) \ = \ \frac{1}{\ell ^{2}}\mathcal{N}^{\prime }(u,\phi )\;,%
\quad\quad\mathcal{\dot{N}}(u,\phi ) \ =  \ \mathcal{M}^{\prime
}(u,\phi )\;,
\label{BMS7}
\end{equation}%
where dot and prime denote derivatives with respect to the coordinates $u $ and $\phi$, respectively, whose solutions are
\begin{equation}\label{Lpm}
\mathcal{M} \ = \ \frac{\mathcal{L}^+ + \mathcal{L}^-}{\ell}\,, \hspace{1cm} \mathcal{N} \ = \  \mathcal{L}^+ - \mathcal{L}^-\;,
\end{equation}
with $\mathcal{L}^\pm = \mathcal{L}^\pm(x^\pm)$ and $x^\pm = \phi \pm \frac{1}{\ell} u$. The field equation associated with the gauge field $\tilde\sigma^a$ leads to the extra condition
\begin{equation}\label{eqr}
\dot{\mathcal{R}}(u,\phi )=\mathcal{N}^{\prime }(u,\phi )\,,
\end{equation}
which yields
\begin{equation}
\mathcal{R}=\ell \, \left(\mathcal{L}^+ + \mathcal{L}^- -2 \mathcal{L}\right)\,,\quad\quad\mathcal{L}=\mathcal{L}(\phi).  \label{BMS8}
\end{equation}
Now, the transformation in \eqref{tilde} induces the following relationship among the gauge fields of the BMS and the original AdSL basis
\begin{equation}
e^{a}=\tilde{e}^{a}\,, \hspace{1cm} \omega ^{a}=\tilde{\sigma}^{a}\,, \hspace{1cm}\sigma ^{a}= \tilde{\omega}^{a}-\tilde{\sigma}^{a} \,.
\end{equation}%
Hence, the solution for AdSL CS gravity is given by 
\begin{eqnarray} \label{bmsg}
e^{0} & = & \dfrac{1}{2\ell^{2}}\left(\ell^{2}\, \mathcal{M}-r^{2}\right) \dd u-\dd r+\dfrac{1}{2}\mathcal{N} \dd \phi\,,\hspace{0.5cm}  e^{1} =\dd u\,, \quad  e^{2}  =r \dd  \phi \,,\\
\omega ^{0}  & = &\dfrac{1}{2\ell^{2}}\left(\ell^{2}\, \mathcal{M}-\mathcal{R}\right) \dd \phi \,,\hspace{3cm}  \omega ^{1}  =\dd \phi \,, \hspace{0.4cm}  \omega ^{2}  =0\,,\\
\kappa ^{0}  &= &-\dfrac{1}{2\ell^2}\mathcal{N} \dd u+\dfrac{1}{2\ell^2}\left(r^{2}-\mathcal{R}\right) \dd \phi \,,\hspace{1.4cm}  \kappa ^{1}  =0\,, \hspace{0.6cm}  \kappa ^{2}  =-\dfrac{r}{\ell^2} \dd u\,.
\end{eqnarray}
 It is important to remark that the integration constant $a$ for this gauge vanishes. Moreover, it is direct to show that enforcing  the functions $\mathcal{M}$, $\mathcal{N}$ and $\mathcal{R} $ to be the constants
\begin{equation}
\mathcal{M}\left( u,\phi \right) =M\,,\hspace{1cm} \mathcal{N}\left(
u,\phi \right) =-J\,,\hspace{1cm} \mathcal{R} \left(u,\phi \right)=\ell ^{2} \left(b+M\right)\,,
\label{BMS13}
\end{equation}
and the arbitrary functions to be fixed by
\begin{equation}
I(r)=r N_{\varphi }\,, \hspace{1cm} H(r)=\frac{r N^2_{\varphi }}{N^{2}} \, + \, \frac{1}{N} \frac{\dd N}{\dd r}\,,\hspace{1cm} F(r)=\frac{b+r^2 N^2_{\varphi}-N^{2}}{2N} \,, \label{gauge_fix}
\end{equation}
we connect the results of the BMS basis \eqref{BMS3} with the ones written in the AdSL basis of the stationary solution \eqref{ADM}. In conclusion, the BMS gauge basis fixes the form of the arbitrary functions $I, H, F$ in terms of the BTZ geometric data. Then, the phase space of this family of solutions is spanned by the integration constants $M, J$ and $b$.

\subsection{Asymptotic dynamics}\label{Section:asymptoticsymmetry}
The thermodynamic properties of a black hole can be ascertained by assessing the Euclidean action associated with the black hole solution. The thermodynamic functions mostly get contributions from the boundary terms. Moreover, the extensive thermodynamic quantities, the global charges defined at infinity, depend on asymptotic symmetries. Therefore, it is essential to understand the behavior of dynamic fields in the asymptotic region. 

In order to calculate the thermodynamics of the BTZ type solution \eqref{ADM}, we have to consider suitable fall-off conditions for the gauge field at infinity. The gauge connection $A$ associated with the gauge fields \eqref{bmsg} written in the $\lbrace J_{a},P_{a},Z_{a}\rbrace$ basis is
\begin{eqnarray}\label{connection}
A &=& \frac{1}{2\ell^2}\, \left( \mathcal{N}\left( u,\phi \right) \dd u+ \left(\mathcal{R}
\left( u,\phi \right)  - r^{2}\right)\dd \phi \right) Z_{0}+\frac{r}{\ell^2} \dd u Z_{2}
\nonumber \\
\nonumber &&+\, \left(  \frac{1}{2\ell^2 }\left(\ell^{2}\mathcal{M}\left( u,\phi \right) -r^{2}\right)\dd u-\dd r+\frac{1}{2}
\mathcal{N}\left( u,\phi \right) \dd \phi \right)
\frac{P_{0}}{\ell} \, + \, \frac{P_{1}}{\ell} \dd u \, + \, r\, \dd \phi  P_{2}   \\
&&+\, \left( \frac{1}{2\ell^{2}}\left(\ell^{2}\, \mathcal{M}\left( u,\phi \right) -\mathcal{R} \left( u,\phi \right) \right) \dd \phi \right) J_{0}+\dd \phi J_{1}\,.
\end{eqnarray}
 It is possible to remove the radial coordinate $r$ by performing an appropriate gauge transformation on \eqref{connection}. Indeed, doing
\begin{equation}
A=\theta^{-1}\dd \theta +\theta^{-1}a\, \theta\,,
\end{equation}%
with the group element $\theta=e^{-\frac{r}{\ell} P_{0}}$. Then, the asymptotic gauge connection is 
\begin{eqnarray}\label{aadslor}
a &=& \frac{1}{2 \ell^2}\, \left( \mathcal{N}\left( u,\phi \right) \dd u \, + \, \mathcal{R}
\left( u,\phi \right) \dd\phi \right) Z_{0} \, + \, \frac{1}{2 \ell}\, \left( \mathcal{M}
\left( u,\phi \right) \dd u+\mathcal{N}\left( u,\phi \right) \dd \phi
\right) P_{0}  \notag \\
&&+\ \frac{\dd u}{\ell}P_{1}+ \frac{1}{2\ell^{2}}\left(\ell^{2}\, \mathcal{M}\left( u,\phi \right) -\mathcal{R} \left( u,\phi \right) \right) \dd \phi  J_{0}+\dd \phi
J_{1}\,.  
\end{eqnarray}
Since we are interested in the thermodynamic properties of the solution \eqref{bmsg}, it is convenient to rewrite \eqref{aadslor} in the basis $\lbrace J_a^{+},J_a^{-},\hat{J}_a\rbrace$, then   
\begin{equation}
a=\left(a^{+}_{u}+a^{-}_{u} \right)\frac{\dd u}{\ell}\ + \ \left(a_\phi^{+}+a_\phi^{-}+\hat{a}_\phi \right)\dd \phi \,,
\end{equation}
where
\begin{eqnarray}
a_\phi^{\pm}&=&\frac{1}{\ell}\mathcal{L}^{\pm}(u,\phi)J_0^{\pm}+J_1^{\pm}\,, \label{Eq:aphipm}\\
\hat{a}_\phi &=& \frac{1}{\ell}\mathcal{L}(\phi) \hat{J}_0+\hat{J}_{1}\,,\label{Eq:aphihat}
\end{eqnarray}
and
\begin{equation}\label{eq:basisLpm}
\mathcal{L}^{\pm}(u,\phi)=\frac{1}{2}\left(\ell \, \mathcal{M}(u,\phi)\,  \pm \, \mathcal{N}(u,\phi)\right)\,,\hspace{1cm} \mathcal{L}= \frac{1}{2\ell}\left(\ell^{2} \, \mathcal{M}(u,\phi) - \mathcal{R}(u,\phi)\right)\,.
\end{equation}
The asymptotic gauge fields $a_{u}^{+}$ and $a_{u}^{-}$ will be specified below. On the other hand, the asymptotic form of the connection  \eqref{connection} is preserved under gauge transformations of the form $\delta_\lambda a=\mathrm{d}\lambda + \left[ a,\lambda \right]$, where $\lambda$ is zero-form a gauge parameter 
\begin{equation}
\Lambda =\theta^{-1}\lambda \theta\,,\qquad \lambda = \frac{1}{\ell}\varepsilon^{a}\left( u,\phi \right) P_{a}+\chi ^{a}\left( u,\phi \right)
J_{a}+\gamma ^{a}\left( u,\phi
\right) Z_{a}\,.  \label{lambda2}
\end{equation}
Note that $\lambda=\lambda(u,\phi)$ depends only on spacetime coordinates $u$ and $\phi$ because we gauged away its $r$-dependence. The functions $\varepsilon^{a}, \chi ^{a}$, and $\gamma ^{a}$ are the symmetry parameters along the generators of the AdSL algebra. Taking the connection \eqref{aadslor} and the $r$-independent gauge parameter \eqref{lambda2}, one can show its components can be solved in terms of three arbitrary functions, i.e. $\lambda= \lambda (Y,f,h)$, with $Y = Y(\phi)$, $f =f(u, \phi)$, and $h=h(u,\phi)$. Explicitly, 
\begin{eqnarray}
\lambda &=&\left( \frac{1}{2\ell^{2}} \left(\ell^{2}\, \mathcal{M}-\mathcal{R}\right)Y-Y^{\prime\prime}\right)J_0+Y J_1-Y^{\prime}J_2+\frac{1}{2\ell}\left(\mathcal{M}f+\mathcal{N}Y+\mathcal{N}h-2f^{\prime\prime} \right) P_0 \nonumber\\
&&+\ \frac{f}{\ell} P_1- \frac{f^{\prime}}{\ell}P_2+ \frac{1}{2\ell^{2}}\left( \ell^{2}\, \mathcal{M}h+\mathcal{N}f+\mathcal{R} Y-2\ell^{2}h^{\prime\prime} \right)Z_0+ h Z_1-h^{\prime}Z_2\,,
\end{eqnarray}
whenever the functions $\mathcal{M},\mathcal{N}$ and $\mathcal{R}$ transform according to~\cite{Concha:2018jjj}
\begin{subequations}
\begin{align}
\delta_{\lambda} \mathcal{M} &=\mathcal{M}^{\prime }Y+2\mathcal{M}Y^{\prime
}-2Y^{\prime \prime \prime }-2 h ^{\prime \prime
\prime }+h\mathcal{M}^{\prime } + 2\mathcal{M}h
^{\prime }+ \frac{2\mathcal{N}f ^{\prime }}{\ell^2}+\frac{f%
\mathcal{N}^{\prime }}{\ell^2}  \,,   \label{deltam} \\
\delta_{\lambda} \mathcal{N} &=\mathcal{M}^{\prime }f+2\mathcal{M}%
f ^{\prime }-2f ^{\prime \prime \prime }+\mathcal{N}%
^{\prime }Y+2\mathcal{N}Y^{\prime }+ 2\mathcal{N}%
h ^{\prime }+h \mathcal{N}^{\prime } \,,
\label{deltan}  \\ \medskip
\delta_{\lambda} \mathcal{R} &= \ell^2 \left(\mathcal{M}^{\prime }h+2\mathcal{M}h
^{\prime }-2h ^{\prime \prime \prime }\right)+\mathcal{N}^{\prime
}f+2\mathcal{N}f ^{\prime }+\mathcal{R} ^{\prime
}Y+2\mathcal{R} Y^{\prime }\,.   \label{deltalpha}
\end{align}
\end{subequations}
We will now determine the asymptotic form of the gauge fields along time evolution. By considering one of the components of the field equations, we find 
\begin{equation}
F_{iu} =0\hspace{0.8cm} \Longrightarrow  \hspace{0.8cm} \dot{A_i}=\partial_i A_u +[A_i ,\, A_u]\,,
\end{equation}
with $i=1,2$ denoting the spatial coordinates. From here, one can recognize the structure of a gauge transformation equation for the gauge field $A_{i}$. Therefore, we can interpret this equation as the time evolution of $A_{i}$ as a gauge transformation generated by the parameter $A_u$. To preserve the asymptotic symmetries, the Lagrange multiplier must be $A_u=\theta^{-1} a_u \theta$, with 
\begin{equation}\label{Eq:auwithchemicals}
a_u=\lambda \left(\mu,\xi,\rho\right)\,,
\end{equation}
where we have included the chemical potentials $\mu$, $\xi$ and $\rho$, following the analysis presented in \cite{Henneaux:2013dra}.  Therefore, the time evolution of the gauge fields in the asymptotic region is given by the following set of differential equations
\begin{subequations}\label{Eq:asympfieldequations}
\begin{eqnarray}
\dot{\mathcal{M}}&=& \mathcal{M}^{\prime}\mu +2\mathcal{M}\mu^{\prime}+ \mathcal{M}^{\prime} \rho+ 2\mathcal{M} \rho^{\prime}+\frac{\mathcal{N}^{\prime} \xi}{\ell^{2}}+\frac{2\mathcal{N} \xi^{\prime}}{\ell^{2}}-2\mu^{\prime \prime \prime}- \rho^{\prime \prime \prime}\,, \\
\dot{\mathcal{N}}&=& \mathcal{M}^{\prime}\xi + 2\mathcal{M}\xi^{\prime}+ \mathcal{N}^{\prime}\mu +2\mathcal{N}\mu^{\prime}+\mathcal{N}^{\prime}\rho+ 2\mathcal{N}\rho^{\prime}-2\xi^{\prime \prime \prime}\,,\\
\dot{\mathcal{R}}&=& \ell^2 \mathcal{M}^{\prime}\rho +2 \ell^2 \mathcal{M}\rho^{\prime} +  \mathcal{N}^{\prime} \xi +2 \mathcal{N} \xi^{\prime}+ \mathcal{R}^{\prime}\mu+ 2\mathcal{R}\mu^{\prime}-2\ell^2 \rho^{\prime \prime \prime}\,.
\end{eqnarray}
\end{subequations}
It is clear from these equations that configurations with constant values of $\mathcal{M},\mathcal{N}$ and $\mathcal{R}$, as well as their corresponding chemical potentials $\mu$, $\xi$ and $\rho$, solve the field equations \eqref{Eq:asympfieldequations}. Considering \eqref{Eq:auwithchemicals}, the gauge connection \eqref{aadslor} takes the form

\begin{equation}\label{aconnec}
a=a_u \dd u \ + \ \left(  \frac{1}{2} \left( \ell^{2}\, \mathcal{M} -\mathcal{R} \right)J_{0}+J_{1}+\frac{\mathcal{N}}{2\ell} P_0 + \frac{\mathcal{R}}{2\ell^2} Z_0 \right) \dd \phi   \,,
\end{equation}
with
\begin{eqnarray}
a_u(\mu,\xi,\rho) &=& \frac{1}{2\ell^{2}} \left(\ell^{2}\, \mathcal{M}-\mathcal{R}\right)\mu J_0+\mu J_1+ \frac{1}{2\ell}\left(\mathcal{M}\xi+\mathcal{N}\mu+\mathcal{N}\rho \right)P_0+\frac{\xi}{\ell} P_1 \nonumber\\
&& +\ \frac{1}{2\ell^{2}} \left( \ell^{2}\mathcal{M}\rho+\mathcal{N}\xi+\mathcal{R} \mu \right)Z_0+ \rho Z_1\,.
\end{eqnarray}
Writing these results above in the basis $\lbrace J_a^{+},J_a^{-},\hat{J}_a\rbrace$, the time evolution of the gauge fields is given by
\begin{eqnarray}
\dot{\mathcal{L}}^{\pm}&=&\mathcal{L}^{\pm \prime}\nu_{\pm}+\mathcal{L}^{\pm \prime} \nu +2\mathcal{L}_{\pm} \nu_{\pm}^{\prime}+2\mathcal{L}_{\pm} \nu^{\prime}-\ell\left(\nu_{\pm}^{\prime \prime \prime}+ \nu^{\prime \prime \prime}\right)\,, \\
\dot{\mathcal{L}}&=& \mathcal{L}^{\prime} \nu +2 \mathcal{L}\nu ^{\prime}-\ell \nu ^{\prime \prime \prime}\,,
\end{eqnarray}
where
\begin{equation}
\nu_{\pm}=\rho \ \pm  \ \frac{\xi}{l}\,, \hspace{1cm} \nu= \mu\,,
\end{equation}
are the chemical potentials in this basis. Finally, in the rest frame, the asymptotic gauge connection \eqref{aconnec} reads
\begin{equation}\label{aconnecJ0}
a= \left(a^{+}_{u} + a^{-}_{u}+\hat{a}_u \right) \dd u \ +\ \left(a_{\phi}^{+}+a_{\phi}^{-}+\hat{a}_{\phi} \right) \dd \phi    \,,
\end{equation}
with
\begin{eqnarray}
a_{u}^{\pm}&=&\pm \left(\nu_{\pm}+ \nu \right) \left(\frac{1}{\ell}\mathcal{L}^{\pm}J_0^{\pm}+J_1^{\pm}\right)\,,\\
\hat{a}_{u} &=&\nu \left(\frac{\mathcal{L}}{\ell} \hat{J}_0+\hat{J}_1  \right)\,,
\end{eqnarray}
where $a_{\phi}^{\pm}$ and $\hat{a}_{\phi}$ are given in Eqs.~\eqref{Eq:aphipm} and \eqref{Eq:aphihat}. In the following section, we will compute the black hole entropy following the known formulae of the three-dimensional CS theory \cite{Bunster:2014mua}.  
\section{Thermodynamics}\label{Section:Thermo}
We now study the thermodynamics of the stationary solution \eqref{bmsg}. For this purpose, we follow for a large part the treatment presented in~\cite{Carlip:1994gc,Henneaux:2013dra,Bunster:2014mua}. In general, black hole entropy is determined by establishing regularity conditions and global charges. In a CS gauge theory of gravity, regularity conditions are achieved by requiring trivial holonomies along contractible cycles. 
%
\subsection{Entropy}
To set regularity conditions easily, we will use the results obtained on the basis $\lbrace J_a^{+},J_a^{-},\hat{J}_a\rbrace$. The connection one-form takes values in the three copies of the Lorentz algebra such that
\begin{equation}
A=A^{+} +A^{-} + \hat{A}\,,
\end{equation}
with
\begin{equation}
A^{\pm}=\left(\omega^{a} \pm \frac{e^{a}}{\ell}+\sigma^{a}\right)J^{\pm}_a\,, \hspace{1cm}\hat{A}=\omega^{a}\hat{J}_a\,.
\end{equation}
For this gauge connection, the CS action \eqref{cs} is rewritten as 
\begin{equation}
I_{\mathrm{CS}}\left[A\right] \ = \ I_{\mathrm{CS}}\left[A^{+}\right]+I_{\mathrm{CS}}\left[A^{-}\right]+I_{\mathrm{CS}}\left[\hat{A}\right]\,,
\end{equation}
where we have considered the following non-vanishing components of the invariant tensor 
\begin{equation} \label{invten3sl}
\left\langle J_{a}^{\pm}, \, J_{b}^{\pm}\right\rangle \, = \, \frac{1}{2}\left( \alpha_{2}\pm \alpha_{1}\right)\eta_{ab}
\,, \hspace{1cm}
\left\langle \hat{J}_{a}, \, \hat{J}_{b}\right\rangle \, = \, \left(\alpha_{0}- \alpha_{2}\right)\eta_{ab}\,.
\end{equation}
In the microcanonical ensemble, the entropy can then be obtained from the following expression \cite{Bunster:2014mua}
\begin{equation}
S=k \,\left( \left\langle a_{u}^{+}\,, a_{\phi}^+\right\rangle +\left\langle a_{u}^{-}\,, a_{\phi}^{-} \right\rangle + \left\langle \hat{a}_{u} \,, \hat{a}_{\phi}\right\rangle\right) \Bigg|_{\text{on-shell}}\,.\label{entropy}
\end{equation}
 Following the nomenclature of \cite{Bunster:2014mua}, without loss of generality, we shall make the assumption that we are in a ``rest frame'' such that the functions $\lbrace \mathcal{L}^\pm,\mathcal{L}\rbrace$ and $\lbrace{\nu^\pm,\nu \rbrace}$ are constants at $r\to \infty$, i.e.
\begin{equation}\label{fieldInf}
\mathcal{L}^{\pm}\left(u,\phi \right) \xrightarrow[r\to \infty]{} \frac{\mathcal{L}_{0}^{\pm}}{2\pi}\,,\hspace{0.5cm} \mathcal{L}(u,\phi) \xrightarrow[r\to \infty]{}  \frac{\mathcal{L}_{0}}{2\pi}\,, \hspace{0.5cm}\nu^{\pm}(u,\phi) \xrightarrow[r\to \infty]{}  \frac{\nu_{0}^{\pm}}{2\pi}\,, \hspace{0.5cm}\nu(u,\phi) \xrightarrow[r\to \infty]{}  \frac{\nu_{0}}{2\pi}\,.
\end{equation} 
In addition, we conveniently rescale the fields $\lbrace \mathcal{L}^\pm,\mathcal{L}\rbrace$, such that
\begin{subequations}\label{Figauge}
\begin{eqnarray}
a_\phi^{\pm}&=&\frac{4\pi}{k}\mathcal{L}^{\pm}J_0^{\pm}+J_1^{\pm}\,,\\
\hat{a}_\phi &=& \frac{4\pi}{k}\mathcal{L} \hat{J}_0+\hat{J}_1\,,\\
a_u^{\pm}&=&\left(\nu_{\pm}+ \nu \right) \left( \frac{4\pi}{k}\mathcal{L}^{\pm}J_0^{\pm}+J_1^{\pm}\right)\,, \\
\hat{a}_u &=&\nu \left(\frac{4\pi}{k}\mathcal{L} \hat{J}_0+\hat{J}_1 \right)\,.
\end{eqnarray}
\end{subequations}
From these results and considering the formulas \eqref{invten3sl}-\eqref{entropy}, we deduce the general expression for the entropy
\begin{equation}
S=4\pi\left(2\alpha_0 \nu \mathcal{L}+\alpha_1 \left( (\nu_{+}+\nu)\mathcal{L}^+ + (\nu_{-}+\nu)\mathcal{L}^{-} \right)+\alpha_2  \left( (\nu_{+}+\nu)\mathcal{L}^{+}-(\nu_{-}+\nu)\mathcal{L}^{-}-2 \nu \mathcal{L} \right) \right)\,.
\end{equation}
In order to fix the chemical potentials, we require the holonomies $H$ along the thermal circle to be trivial, namely
\begin{equation}\label{Holo}
H=e^{\mathrm{i} a_{u}(r_{+})}\Bigg|_{\text{on-shell}}=-\mathbb{1}\,,
\end{equation}
where $\mathbb{1}$ is the identity generator of the AdSL algebra and $r_{+}$ is the largest root of $N(r)=0$. Then, for each temporal gauge connection \eqref{Figauge} satisfying the regularity condition \eqref{Holo}, we obtain a set of equations that allows us to find $\nu_{\pm}$ and $\nu$. According to this condition, the eingenvalues of $\mathrm{i}a_u$ are given by $\pm \, \mathrm{i} \pi$. Therefore,
\begin{subequations}\label{Treq}
\begin{eqnarray}
\text{tr}\left[\left(a_{u}^{+}\right)^2\right]&=& 2\pi^2\,,\\
\text{tr}\left[\left(a_{u}^{-}\right)^2\right]&=& 2\pi^2\,,\\
\text{tr}\left[\left(\hat{a}_{u}\right)^2\right]&=& 2\pi^2\,.
\end{eqnarray}
\end{subequations}
On the other hand, because of the isomorphism $\mathfrak{so}(2,1)\simeq \mathfrak{sl}(2,\mathbb{R})$, then we conveniently choose 
\begin{equation}
\hat{J}_0^{\pm} \simeq j_0\,, \hspace{1cm} \hat{J}_1^{\pm} \simeq  j_1\,,
\end{equation}
where $j_{0}$ and $j_{1}$ are both $\mathfrak{sl} \left(2,\mathbb{R} \right)$ elements.\footnote{See, for instance, appendix of \cite{Donnay:2016iyk}} This isomorphism allows us to use the $\mathfrak{sl} \left(2,\mathbb{R} \right)$ matrix representation to compute the holonomy \eqref{Treq}.  
Therefore, from \eqref{Treq}, $\nu_{\pm}$ and $\nu$ are given by
\begin{equation}\label{ChePote}
\nu_{+}+\nu = \sqrt{\frac{\pi k}{2 \mathcal{L}^+}}\,, \hspace{1cm}
\nu_{-}+\nu = \sqrt{\frac{\pi k}{2 \mathcal{L}^-}}\,, \hspace{1cm}
\nu = \sqrt{\frac{\pi k}{ 2 \mathcal{L}}}\,.
\end{equation}
Then, by considering \eqref{fieldInf} and \eqref{ChePote} the entropy \eqref{entropy} turns out to be
\begin{equation}
S=2\pi \sqrt{k}\, \left( \alpha_1 \left(\sqrt{\mathcal{L}^{+}_{0}}+ \sqrt{\mathcal{L}^{-}_{0}} \right)+\alpha_2 \left(\sqrt{\mathcal{L}^{+}_{0}}-\sqrt{\mathcal{L}^{-}_{0}}\right) +2 \left( \alpha_{0} - \alpha_{2}\right)\sqrt{\mathcal{L}_{0}}\right)\,.\label{entr0}
\end{equation}
From the expression \eqref{eq:basisLpm} we can rewrite (\ref{entr0}) in terms of the integration constants $M$, $J$ and $b$ as 
\begin{equation}\label{eq:entropyLorentzbasis}
 S= \pi \sqrt{2k}\left( \alpha_1\left( \sqrt{\ell M-J} - \sqrt{\ell M+J} \right)  +\ \alpha_2 \left(  \sqrt{\ell M-J} + \sqrt{\ell M+J}\right) +4 \sqrt{-\ell b} \left(\alpha_0-\alpha_{2} \right)  \right)\,.
\end{equation}
Notice that the term proportional to $\alpha_{2}$ gives the standard result of the BTZ black hole. However, we now have the contribution of the exotic term $\alpha_{1}$ and also from the integration constant $b$. At this level, we cannot set $\alpha_{0}-\alpha_{2} =0$ because the bilinear form \eqref{Eq:AdSLorentzpairing} degenerates. One can also write the entropy in terms of the global charges associated with the isometries of this new geometry. To this end, in the next section, we compute the global charges and show the appearance of a new conserved charge.
\subsection{Global and gauge charges}\label{Seciton:globalchargesandentropy}
%
Let us consider diffeomorphisms and gauge symmetries together, both parameterized by $\epsilon= (\xi ,\lambda)$, with $\xi$ a vector field and $\lambda$ a Lie algebra valued gauge parameter. Thus, a general infinitesimal symmetry transformation reads\footnote{We use the notation $\xi\inter = i_\xi$ for the interior product, for instance $\xi\inter e^a=\xi^\mu e\indices{^a_\mu}$.}
\begin{equation}\label{Eq:symmetryCS}
\delta A_{\epsilon} = \mathcal{L}_{\xi} A-\mathrm{D}_{A}\lambda^{\prime} = \xi \inter F-\mathrm{D}_{A}\lambda \,,
\end{equation} 
where we have defined the exterior covariant derivative $\mathrm{D}_{A} (\cdot) \equiv \dd(\cdot) +[A,\, (\cdot)]$ and the displaced parameter $\lambda= \lambda^{\prime}- \xi \inter A$. Following \cite{Frodden:2019ylc}, to define a conserved charge, we demand the so-called exact symmetry condition, which consists in standing the infinitesimal gauge transformations \eqref{Eq:symmetryCS} to zero and solve for the parameters $\epsilon=(\xi, \lambda)$. For instance, the global charges associated with the invariance under time displacements (mass) and rotations (angular momenta) are computed using, respectively, the Killing vectors $\xi^{\mu}_{(u)}=(1,0,0)$ and $\xi^{\mu}_{(\phi)}=(0,0,-1)$, respectively. The formula to derive these charges, with $u=const$, is given by 
\begin{equation}\label{eq:chargeformula}
\delta Q\left[\lambda \right] =- \frac{k}{2\pi}\, \int \hspace{-0.1cm}\dd\phi \, \left\langle \lambda , \delta a \right\rangle\,.
\end{equation}
where we considered the expression for the asymptotic gauge field $a$ \eqref{aadslor}.\footnote{It is worth noting here that the quasi-local treatment for the charge conservation ensures, given a spacetime with exact symmetries, the independence of the radius $r$. } By choosing $\lambda= - \xi \inter a $, the exact symmetry condition $\delta_{\epsilon}A=0$ is solved for the Killing vectors  $\xi^{\mu}_{(u)}$ and $\xi^{\mu}_{(\phi)}$, only. The charges associated with these isometries become 
\begin{eqnarray}
\delta Q\left[\xi^{\mu}_{(u)} \right] & \equiv& \delta m = \frac{k}{2\ell} \left( \alpha_1 \delta M - \frac{\alpha_2}{\ell} \delta J \right)    \,, \\ 
\delta Q\left[\xi^{\mu}_{(\phi)} \right] & \equiv& \delta j =  \frac{k}{2\ell} \left(   \left(\alpha_{0}-\alpha_{2} \right)\ell \, \delta b + \alpha_1 \delta J - \alpha_{2} \ell \delta M \right) \,.
\end{eqnarray}
The integration along $\phi$ is trivial because the charges are coordinate independent. Moreover, since the phase space variation is linear, this integration is trivial as well. Setting $k\equiv 2\ell$, we get the finite charges
\begin{eqnarray}
 m & = & \alpha_1  M -\frac{\alpha_2}{\ell} J     \,, \\ 
 j &= &      \left(\alpha_{0}-\alpha_{2} \right)\, \ell b+ \alpha_1  J - \alpha_2 \ell \,M\,.
\end{eqnarray}
Notice that the rotating BTZ case corresponds to $(\alpha_{0},\alpha_{1},\alpha_{2}) = (0,1,0)$, and the case $(\alpha_{0},\alpha_{1},\alpha_{2}) = (0,0,-1)$ with $b=0$, corresponds to the exotic BTZ charges, where a reversion of the mass and angular momentum is observed.
Therefore, here we find that global charges essentially generalize the exotic case found by Townsend and Zhang \cite{Townsend:2013ela} and contain, in addition, a torsional correction proportional to $b$ in $j$. 

There is an extra charge associated with a gauge symmetry in this solution. This charge is computed using the gauge parameter $\lambda$ satisfying the exact symmetry condition for $\epsilon=(0,\lambda)$, so that
\begin{equation}
\delta_{\lambda}a=\dd\lambda+ \left[a,\lambda \right] =0\,,
\end{equation}
where $\lambda$ is the algebra valued parameter given in \eqref{lambda2} and $a$ the gauge connection appearing in \eqref{aadslor}. This exactness condition admits the solution for $\varepsilon^a =0$ and $\chi^{a} = - \gamma^{a}$, we get
\begin{equation}\label{eq:torsiongaugeparameter}
\lambda_{\scriptscriptstyle w} \equiv -\frac{b}{2}J_0   + J_1+ \frac{b}{2}Z_0  - Z_1 \,.
\end{equation}
Then, replacing \eqref{eq:torsiongaugeparameter} into \eqref{eq:chargeformula}, and integrating both in $\phi$ spacetime coordinate and in phase space, we obtain the new gauge charge
\begin{equation}\label{eq:newtorsioncharge}
 Q\left[\lambda_{\scriptscriptstyle w}  \right]  \equiv  w =  \left(\alpha_0 -\alpha_2\right) \, \ell b\,.
\end{equation}
This is a novel result of this work. Although AdSL symmetry has three Casimir operators \cite{Diaz:2012zza}, and this new charge is expected, the surprising result here is how the torsion supports it. It could be interesting to study the possible relationship between the constant electromagnetic field of the AdSL particle analysis \cite{Bacry:1970aa,Gomis:2009dm} and this new charge. We leave this for a future work.

From these charges in hand, it is possible to write the entropy \eqref{eq:entropyLorentzbasis} in terms of the global charges $m, j$ and $w$. We find
\begin{equation}
S= 2 \pi  \sqrt{\ell} \left( \sqrt{(\ell m+j-w) (\alpha_1-\alpha_2)} + \sqrt{(\ell m-j+w)(\alpha_1+\alpha_2)} + 4\sqrt{w(\alpha_{2}-\alpha_{0})} \right)\,,
\end{equation}
or in terms of the inner/outer horizons $r_{\pm}$ and the integration constant $b$ becomes
\begin{equation}
S\, = \, 4\pi \left( \alpha_1 r_{+} -\alpha_{2} r_{-} + 2\left(\alpha_{0}-\alpha_{2}\right) \,\ell \sqrt{-b}  \right)\,,
\end{equation}
noticing that the integration constant $b<0$ holds. Here we have assumed that $\alpha_{0}>\alpha_{2}$. The case $\alpha_{0}<\alpha_{2}$ is permissible as well, giving a minus sign in front of the last term.  To recover the result found in \cite{Townsend:2013ela}, one should set $\alpha_{2}\rightarrow - \tilde{\alpha}_{2}$, with $\tilde{\alpha}_{2}>0$ and $b=0$.  

This expression for the entropy differs from the standard and exotic formula because of the term proportional to the new gauge charge $w$. Due to that, this charge is related to the integration constant appearing in the solution for $\sigma^{a}=-\kappa^{a}$ gauge field, this result indicates that spacetime torsion modifies both the black hole entropy and the first law of thermodynamics, as we will see below.

\subsection{The first law and Smarr formula}
%
By studying the thermodynamics of the black hole with torsion \eqref{BMS13}, calculation of the entropy formula \eqref{eq:entropyLorentzbasis} revealed the existence of a new extensive variable associated with the integration constant $b$. 
Therefore, we can establish the first law of black hole mechanics for this family of black hole solution  spanned by the parameters $M, J$, and $b$. With all the global charges at hand,  the first law of thermodynamics reads
\begin{equation}
\delta m=T \delta S+ \Omega\,  \delta j +\mu \delta w \,,
\end{equation}
where we have used the standard formulas to compute the extensive quantities 
\begin{eqnarray}
T& = & \frac{(r_{+}^{2} -r_{-}^{2})}{2\pi \ell^2\, r_{+}}\,, \\
\Omega& =& \frac{r_{-}}{\ell r_{+}}  \,, \\
\mu & =& \frac{2(r_{+}^2 -r_{-}^{2}) -  \sqrt{|b|}\ell r_{-}}{\sqrt{|b|} \ell^2 \, r_{+}}= \frac{4\pi T}{\sqrt{|b|}}-\Omega\,.
\end{eqnarray}
All of these quantities might be further identified with standard physical ones: a temperature (Hawking thermal radiation) $T$, angular velocity of the horizon $\Omega$, and a chemical potential $\mu$ associated with the gauge charge $w$.

Finally, by considering the fact that the entropy is a homogeneous thermodynamical function of degree $1/2$ for $m, j$ and $w$, one easily obtains the Smarr formula for this black solution 
\begin{equation}
S= \frac{2}{T}\left( m- \Omega j - \mu w \right)\,.
\end{equation}
When $w=0$, this expression reduces to the Smarr formula for the BTZ black hole solution. Lastly, using these state variables and the thermodynamic functions shown, it is possible to find other quantities, such as the heat capacity and fluctuation modes, that will be needed, for example, to study the critical behavior and phase transition of this torsional black hole.

\section{Conclusions}\label{Section:conclusions} 
In this paper, we have examined the thermodynamics of BTZ-type black holes in the presence of spacetime torsion. After a suitable gauge fixing, we show that the solution is characterized by three integration constants:  $M, J$ as in standard GR plus a new constant $b$ related to torsion. The presence of this constant generalizes the \textquotedblleft{}exotic'' thermodynamics analysis of Townsend and Zhang by including spacetime torsion in both the global charge formulas and entropy. 
The solution has been obtained from a CS action valued on a semi-simple enlargement of the Poincaré gauge symmetry, namely the AdSL symmetry. This enlargement includes new gauge fields that allow the notion of a torsion different from zero. Further, we found a novel conserved quantity related to torsion. This implies the presence of this new gauge charge in the laws of black hole thermodynamics.  

Investigating the thermodynamics in the $\ell\rightarrow\infty$ limit of the torsional black hole would be interesting. This observation is motivated by taking the flat limit of the stationary solutions in AdSL gravity, arriving at the cosmological solution of Maxwell CS theory~\cite{Concha:2018zeb}. Moreover, it is also interesting to analyze whether it is possible to recover the entropy formula obtained through a microscopic description from the Cardy formula. This is a work in progress. Including matter fields in the AdSL CS theory is another important topic to discuss. Since fermionic fields interact with torsion, this might be the most intriguing scenario to investigate. A more comprehensive interpretation of the conserved torsional charge may be attainable by doing so. Finally, thermodynamic phenomena such as phase transitions and critical behavior are interesting additional topics to investigate in this torsional black hole. These phenomena involving a conserved torsional charge have not been described previously.

\section*{Acknowledgments}
We thank Eloy Ayón-Beato, Fabrizio Canfora, Cristóbal Corral, Branislav Cvetkovic, Ernesto Frodden, Oscar Fuentealba, Joaquim Gomis, Fernando Izaurieta, José Martín-García, Niels Obers, Alfredo Pérez, Lárus Thorlacius, and Jorge Zanelli, for insightful discussions and comments. DH is supported by the Icelandic Research Fund via the Grant of Excellence titled ``Quantum Fields and Quantum Geometry'' and by the University of Iceland Research Fund. L.A. is supported by Fondecyt grants 3220805. OV acknowledges to ICE-CSIC and Fondecyt 11200742.

\bibliography{BHT.bib}

\providecommand{\href}[2]{#2}\begingroup\raggedright\begin{thebibliography}{10}

\bibitem{PhysRevD.7.2333}
J.~D. Bekenstein, \emph{Black holes and entropy},
  \href{https://doi.org/10.1103/PhysRevD.7.2333}{\emph{Phys. Rev. D} {\bfseries
  7} (1973) 2333}.

\bibitem{Hawking:1972aa}
S.~W. Hawking, \emph{Black holes in general relativity},
  \href{https://doi.org/10.1007/BF01877517}{\emph{Communications in
  Mathematical Physics} {\bfseries 25} (1972) 152}.

\bibitem{Bardeen:1973aa}
J.~M. Bardeen, B.~Carter and S.~W. Hawking, \emph{The four laws of black hole
  mechanics}, \href{https://doi.org/10.1007/BF01645742}{\emph{Communications in
  Mathematical Physics} {\bfseries 31} (1973) 161}.

\bibitem{Jacobson:1995ab}
T.~Jacobson, \emph{{Thermodynamics of space-time: The Einstein equation of
  state}}, \href{https://doi.org/10.1103/PhysRevLett.75.1260}{\emph{Phys. Rev.
  Lett.} {\bfseries 75} (1995) 1260}
  [\href{https://arxiv.org/abs/gr-qc/9504004}{{\ttfamily gr-qc/9504004}}].

\bibitem{Padmanabhan:2009vy}
T.~Padmanabhan, \emph{{Thermodynamical Aspects of Gravity: New insights}},
  \href{https://doi.org/10.1088/0034-4885/73/4/046901}{\emph{Rept. Prog. Phys.}
  {\bfseries 73} (2010) 046901}
  [\href{https://arxiv.org/abs/0911.5004}{{\ttfamily 0911.5004}}].

\bibitem{Wald:1993nt}
R.~M. Wald, \emph{{Black hole entropy is the Noether charge}},
  \href{https://doi.org/10.1103/PhysRevD.48.R3427}{\emph{Phys. Rev. D}
  {\bfseries 48} (1993) R3427}
  [\href{https://arxiv.org/abs/gr-qc/9307038}{{\ttfamily gr-qc/9307038}}].

\bibitem{Hawking:1983aa}
S.~W. Hawking and D.~N. Page, \emph{Thermodynamics of black holes in anti-de
  sitter space}, \href{https://doi.org/10.1007/BF01208266}{\emph{Communications
  in Mathematical Physics} {\bfseries 87} (1983) 577}.

\bibitem{Banados:1992wn}
M.~Banados, C.~Teitelboim and J.~Zanelli, \emph{{The Black hole in
  three-dimensional space-time}},
  \href{https://doi.org/10.1103/PhysRevLett.69.1849}{\emph{Phys. Rev. Lett.}
  {\bfseries 69} (1992) 1849}
  [\href{https://arxiv.org/abs/hep-th/9204099}{{\ttfamily hep-th/9204099}}].

\bibitem{Banados:1992gq}
M.~Banados, M.~Henneaux, C.~Teitelboim and J.~Zanelli, \emph{{Geometry of the
  (2+1) black hole}},
  \href{https://doi.org/10.1103/PhysRevD.48.1506}{\emph{Phys. Rev. D}
  {\bfseries 48} (1993) 1506}
  [\href{https://arxiv.org/abs/gr-qc/9302012}{{\ttfamily gr-qc/9302012}}].

\bibitem{Carlip:1995qv}
S.~Carlip, \emph{{The (2+1)-Dimensional black hole}},
  \href{https://doi.org/10.1088/0264-9381/12/12/005}{\emph{Class. Quant. Grav.}
  {\bfseries 12} (1995) 2853}
  [\href{https://arxiv.org/abs/gr-qc/9506079}{{\ttfamily gr-qc/9506079}}].

\bibitem{Martinez:1996gn}
C.~Martinez and J.~Zanelli, \emph{{Conformally dressed black hole in
  (2+1)-dimensions}},
  \href{https://doi.org/10.1103/PhysRevD.54.3830}{\emph{Phys. Rev. D}
  {\bfseries 54} (1996) 3830}
  [\href{https://arxiv.org/abs/gr-qc/9604021}{{\ttfamily gr-qc/9604021}}].

\bibitem{Carlip:1998wz}
S.~Carlip, \emph{{Black hole entropy from conformal field theory in any
  dimension}}, \href{https://doi.org/10.1103/PhysRevLett.82.2828}{\emph{Phys.
  Rev. Lett.} {\bfseries 82} (1999) 2828}
  [\href{https://arxiv.org/abs/hep-th/9812013}{{\ttfamily hep-th/9812013}}].

\bibitem{Cardenas:2014kaa}
M.~Cardenas, O.~Fuentealba and C.~Mart\'\i{}nez, \emph{{Three-dimensional black
  holes with conformally coupled scalar and gauge fields}},
  \href{https://doi.org/10.1103/PhysRevD.90.124072}{\emph{Phys. Rev. D}
  {\bfseries 90} (2014) 124072}
  [\href{https://arxiv.org/abs/1408.1401}{{\ttfamily 1408.1401}}].

\bibitem{Perez:2015kea}
A.~Perez, M.~Riquelme, D.~Tempo and R.~Troncoso, \emph{{Conserved charges and
  black holes in the Einstein-Maxwell theory on AdS$_{3}$ reconsidered}},
  \href{https://doi.org/10.1007/JHEP10(2015)161}{\emph{JHEP} {\bfseries 10}
  (2015) 161} [\href{https://arxiv.org/abs/1509.01750}{{\ttfamily
  1509.01750}}].

\bibitem{Erices:2017nta}
C.~Erices, O.~Fuentealba and M.~Riquelme, \emph{{Electrically charged black
  hole on AdS$_3$ : Scale invariance and the Smarr formula}},
  \href{https://doi.org/10.1103/PhysRevD.97.024037}{\emph{Phys. Rev. D}
  {\bfseries 97} (2018) 024037}
  [\href{https://arxiv.org/abs/1710.05962}{{\ttfamily 1710.05962}}].

\bibitem{ACHUCARRO198689}
A.~Ach{\'u}carro and P.~Townsend, \emph{A chern-simons action for
  three-dimensional anti-de sitter supergravity theories},
  \href{https://doi.org/https://doi.org/10.1016/0370-2693(86)90140-1}{\emph{Physics
  Letters B} {\bfseries 180} (1986) 89}.

\bibitem{WITTEN198846}
E.~Witten, \emph{2 + 1 dimensional gravity as an exactly soluble system},
  \href{https://doi.org/https://doi.org/10.1016/0550-3213(88)90143-5}{\emph{Nuclear
  Physics B} {\bfseries 311} (1988) 46}.

\bibitem{RevModPhys.48.393}
F.~W. Hehl, P.~von~der Heyde, G.~D. Kerlick and J.~M. Nester, \emph{General
  relativity with spin and torsion: Foundations and prospects},
  \href{https://doi.org/10.1103/RevModPhys.48.393}{\emph{Rev. Mod. Phys.}
  {\bfseries 48} (1976) 393}.

\bibitem{Hehl:1994ue}
F.~W. Hehl, J.~D. McCrea, E.~W. Mielke and Y.~Ne'eman, \emph{{Metric affine
  gauge theory of gravity: Field equations, Noether identities, world spinors,
  and breaking of dilation invariance}},
  \href{https://doi.org/10.1016/0370-1573(94)00111-F}{\emph{Phys. Rept.}
  {\bfseries 258} (1995) 1}
  [\href{https://arxiv.org/abs/gr-qc/9402012}{{\ttfamily gr-qc/9402012}}].

\bibitem{Blagojevic:2013xpa}
M.~Blagojevi{\'c} and F.~W. Hehl, eds., \emph{{Gauge Theories of Gravitation}}.
  World Scientific, Singapore, 2013.

\bibitem{MIELKE1991399}
E.~W. Mielke and P.~Baekler, \emph{Topological gauge model of gravity with
  torsion},
  \href{https://doi.org/https://doi.org/10.1016/0375-9601(91)90715-K}{\emph{Physics
  Letters A} {\bfseries 156} (1991) 399}.

\bibitem{Garcia:2003nm}
A.~A. Garcia, F.~W. Hehl, C.~Heinicke and A.~Macias, \emph{{Exact vacuum
  solution of a (1+2)-dimensional Poincare gauge theory: BTZ solution with
  torsion}}, \href{https://doi.org/10.1103/PhysRevD.67.124016}{\emph{Phys. Rev.
  D} {\bfseries 67} (2003) 124016}
  [\href{https://arxiv.org/abs/gr-qc/0302097}{{\ttfamily gr-qc/0302097}}].

\bibitem{Blagojevic:2006jk}
M.~Blagojevic and B.~Cvetkovic, \emph{{Black hole entropy in 3-D gravity with
  torsion}}, \href{https://doi.org/10.1088/0264-9381/23/14/013}{\emph{Class.
  Quant. Grav.} {\bfseries 23} (2006) 4781}
  [\href{https://arxiv.org/abs/gr-qc/0601006}{{\ttfamily gr-qc/0601006}}].

\bibitem{Blagojevic:2003uc}
M.~Blagojevic and M.~Vasilic, \emph{{Asymptotic symmetries in 3-d gravity with
  torsion}}, \href{https://doi.org/10.1103/PhysRevD.67.084032}{\emph{Phys. Rev.
  D} {\bfseries 67} (2003) 084032}
  [\href{https://arxiv.org/abs/gr-qc/0301051}{{\ttfamily gr-qc/0301051}}].

\bibitem{Blagojevic:2006hh}
M.~Blagojevic and B.~Cvetkovic, \emph{{Black hole entropy from the boundary
  conformal structure in 3D gravity with torsion}},
  \href{https://doi.org/10.1088/1126-6708/2006/10/005}{\emph{JHEP} {\bfseries
  10} (2006) 005} [\href{https://arxiv.org/abs/gr-qc/0606086}{{\ttfamily
  gr-qc/0606086}}].

\bibitem{Hoseinzadeh:2014bla}
S.~Hoseinzadeh and A.~Rezaei-Aghdam, \emph{{(2$+$1)-dimensional gravity from
  Maxwell and semisimple extension of the Poincar{\'e} gauge symmetric
  models}}, \href{https://doi.org/10.1103/PhysRevD.90.084008}{\emph{Phys. Rev.}
  {\bfseries D90} (2014) 084008}
  [\href{https://arxiv.org/abs/1402.0320}{{\ttfamily 1402.0320}}].

\bibitem{Soroka:2006aj}
D.~V. Soroka and V.~A. Soroka, \emph{{Semi-simple extension of the
  (super)Poincare algebra}},
  \href{https://doi.org/10.1155/2009/234147}{\emph{Adv. High Energy Phys.}
  {\bfseries 2009} (2009) 234147}
  [\href{https://arxiv.org/abs/hep-th/0605251}{{\ttfamily hep-th/0605251}}].

\bibitem{Izaurieta:2006zz}
F.~Izaurieta, E.~Rodriguez and P.~Salgado, \emph{{Expanding Lie (super)algebras
  through Abelian semigroups}},
  \href{https://doi.org/10.1063/1.2390659}{\emph{J. Math. Phys.} {\bfseries 47}
  (2006) 123512} [\href{https://arxiv.org/abs/hep-th/0606215}{{\ttfamily
  hep-th/0606215}}].

\bibitem{Diaz:2012zza}
J.~Diaz, O.~Fierro, F.~Izaurieta, N.~Merino, E.~Rodriguez, P.~Salgado et~al.,
  \emph{{A generalized action for (2 + 1)-dimensional Chern-Simons gravity}},
  \href{https://doi.org/10.1088/1751-8113/45/25/255207}{\emph{J. Phys. A}
  {\bfseries 45} (2012) 255207}
  [\href{https://arxiv.org/abs/1311.2215}{{\ttfamily 1311.2215}}].

\bibitem{Fierro:2014lka}
O.~Fierro, F.~Izaurieta, P.~Salgado and O.~Valdivia, \emph{{Minimal AdS-Lorentz
  supergravity in three-dimensions}},
  \href{https://doi.org/10.1016/j.physletb.2018.10.066}{\emph{Phys. Lett. B}
  {\bfseries 788} (2019) 198}
  [\href{https://arxiv.org/abs/1401.3697}{{\ttfamily 1401.3697}}].

\bibitem{Gomis:2009dm}
J.~Gomis, K.~Kamimura and J.~Lukierski, \emph{{Deformations of Maxwell algebra
  and their Dynamical Realizations}},
  \href{https://doi.org/10.1088/1126-6708/2009/08/039}{\emph{JHEP} {\bfseries
  08} (2009) 039} [\href{https://arxiv.org/abs/0906.4464}{{\ttfamily
  0906.4464}}].

\bibitem{Schrader:1972zd}
R.~Schrader, \emph{{The maxwell group and the quantum theory of particles in
  classical homogeneous electromagnetic fields}},
  \href{https://doi.org/10.1002/prop.19720201202}{\emph{Fortsch. Phys.}
  {\bfseries 20} (1972) 701}.

\bibitem{Blencowe:1988gj}
M.~P. Blencowe, \emph{{A Consistent Interacting Massless Higher Spin Field
  Theory in $D$ = (2+1)}},
  \href{https://doi.org/10.1088/0264-9381/6/4/005}{\emph{Class. Quant. Grav.}
  {\bfseries 6} (1989) 443}.

\bibitem{Bergshoeff:1989ns}
E.~Bergshoeff, M.~P. Blencowe and K.~S. Stelle, \emph{{Area Preserving
  Diffeomorphisms and Higher Spin Algebra}},
  \href{https://doi.org/10.1007/BF02108779}{\emph{Commun. Math. Phys.}
  {\bfseries 128} (1990) 213}.

\bibitem{Vasiliev:1995dn}
M.~A. Vasiliev, \emph{{Higher spin gauge theories in four-dimensions,
  three-dimensions, and two-dimensions}},
  \href{https://doi.org/10.1142/S0218271896000473}{\emph{Int. J. Mod. Phys. D}
  {\bfseries 5} (1996) 763}
  [\href{https://arxiv.org/abs/hep-th/9611024}{{\ttfamily hep-th/9611024}}].

\bibitem{Perez:2012cf}
A.~Perez, D.~Tempo and R.~Troncoso, \emph{{Higher spin gravity in 3D: Black
  holes, global charges and thermodynamics}},
  \href{https://doi.org/10.1016/j.physletb.2013.08.038}{\emph{Phys. Lett. B}
  {\bfseries 726} (2013) 444}
  [\href{https://arxiv.org/abs/1207.2844}{{\ttfamily 1207.2844}}].

\bibitem{Salgado:2014qqa}
P.~Salgado and S.~Salgado, \emph{{$\mathfrak{so}(D-1,1)\otimes
  \mathfrak{so}(D-1,2)$ algebras and gravity}},
  \href{https://doi.org/10.1016/j.physletb.2013.11.009}{\emph{Phys. Lett.}
  {\bfseries B728} (2014) 5}.

\bibitem{Townsend:2013ela}
P.~K. Townsend and B.~Zhang, \emph{{Thermodynamics of
  \textquotedblleft{}Exotic\textquotedblright{} Ba\~nados-Teitelboim-Zanelli
  Black Holes}},
  \href{https://doi.org/10.1103/PhysRevLett.110.241302}{\emph{Phys. Rev. Lett.}
  {\bfseries 110} (2013) 241302}
  [\href{https://arxiv.org/abs/1302.3874}{{\ttfamily 1302.3874}}].

\bibitem{Carlip:1994gc}
S.~Carlip and C.~Teitelboim, \emph{{Aspects of black hole quantum mechanics and
  thermodynamics in (2+1)-dimensions}},
  \href{https://doi.org/10.1103/PhysRevD.51.622}{\emph{Phys. Rev. D} {\bfseries
  51} (1995) 622} [\href{https://arxiv.org/abs/gr-qc/9405070}{{\ttfamily
  gr-qc/9405070}}].

\bibitem{Henneaux:2013dra}
M.~Henneaux, A.~Perez, D.~Tempo and R.~Troncoso, \emph{{Chemical potentials in
  three-dimensional higher spin anti-de Sitter gravity}},
  \href{https://doi.org/10.1007/JHEP12(2013)048}{\emph{JHEP} {\bfseries 12}
  (2013) 048} [\href{https://arxiv.org/abs/1309.4362}{{\ttfamily 1309.4362}}].

\bibitem{Bunster:2014mua}
C.~Bunster, M.~Henneaux, A.~Perez, D.~Tempo and R.~Troncoso, \emph{{Generalized
  Black Holes in Three-dimensional Spacetime}},
  \href{https://doi.org/10.1007/JHEP05(2014)031}{\emph{JHEP} {\bfseries 05}
  (2014) 031} [\href{https://arxiv.org/abs/1404.3305}{{\ttfamily 1404.3305}}].

\bibitem{Bacry:1970aa}
H.~Bacry, P.~Combe and J.~L. Richard, \emph{Group-theoretical analysis of
  elementary particles in an external electromagnetic field},
  \href{https://doi.org/10.1007/BF02725178}{\emph{Il Nuovo Cimento A
  (1965-1970)} {\bfseries 67} (1970) 267}.

\bibitem{DESER1982372}
S.~Deser, R.~Jackiw and S.~Templeton, \emph{Topologically massive gauge
  theories},
  \href{https://doi.org/https://doi.org/10.1016/0003-4916(82)90164-6}{\emph{Annals
  of Physics} {\bfseries 140} (1982) 372}.

\bibitem{Nieh:1981ww}
H.~T. Nieh and M.~L. Yan, \emph{{An Identity in Riemann-cartan Geometry}},
  \href{https://doi.org/10.1063/1.525379}{\emph{J. Math. Phys.} {\bfseries 23}
  (1982) 373}.

\bibitem{Chandia:1997jf}
O.~Chandia and J.~Zanelli, \emph{{Torsional topological invariants (and their
  relevance for real life)}}, \href{https://doi.org/10.1063/1.54694}{\emph{AIP
  Conf. Proc.} {\bfseries 419} (1998) 251}
  [\href{https://arxiv.org/abs/hep-th/9708138}{{\ttfamily hep-th/9708138}}].

\bibitem{Concha:2018jjj}
P.~Concha, N.~Merino, E.~Rodr\'\i{}guez, P.~Salgado-Rebolledo and O.~Valdivia,
  \emph{{Semi-simple enlargement of the $\mathfrak{bms}_3$ algebra from a
  $\mathfrak{so}(2,2)\oplus\mathfrak{so}(2,1)$ Chern-Simons theory}},
  \href{https://doi.org/10.1007/JHEP02(2019)002}{\emph{JHEP} {\bfseries 02}
  (2019) 002} [\href{https://arxiv.org/abs/1810.12256}{{\ttfamily
  1810.12256}}].

\bibitem{Barnich:2013yka}
G.~Barnich and H.~A. Gonzalez, \emph{{Dual dynamics of three dimensional
  asymptotically flat Einstein gravity at null infinity}},
  \href{https://doi.org/10.1007/JHEP05(2013)016}{\emph{JHEP} {\bfseries 05}
  (2013) 016} [\href{https://arxiv.org/abs/1303.1075}{{\ttfamily 1303.1075}}].

\bibitem{Barnich:2012aw}
G.~Barnich, A.~Gomberoff and H.~A. Gonzalez, \emph{{The Flat limit of three
  dimensional asymptotically anti-de Sitter spacetimes}},
  \href{https://doi.org/10.1103/PhysRevD.86.024020}{\emph{Phys. Rev.}
  {\bfseries D86} (2012) 024020}
  [\href{https://arxiv.org/abs/1204.3288}{{\ttfamily 1204.3288}}].

\bibitem{Donnay:2016iyk}
L.~Donnay, \emph{{Asymptotic dynamics of three-dimensional gravity}},
  \href{https://doi.org/10.22323/1.271.0001}{\emph{PoS} {\bfseries Modave2015}
  (2016) 001} [\href{https://arxiv.org/abs/1602.09021}{{\ttfamily
  1602.09021}}].

\bibitem{Frodden:2019ylc}
E.~Frodden and D.~Hidalgo, \emph{{Surface Charges Toolkit for Gravity}},
  \href{https://doi.org/10.1142/S0218271820500406}{\emph{Int. J. Mod. Phys. D}
  {\bfseries 29} (2020) 2050040}
  [\href{https://arxiv.org/abs/1911.07264}{{\ttfamily 1911.07264}}].

\bibitem{Concha:2018zeb}
P.~Concha, N.~Merino, O.~Miskovic, E.~Rodr\'\i{}guez, P.~Salgado-Rebolledo and
  O.~Valdivia, \emph{{Asymptotic symmetries of three-dimensional Chern-Simons
  gravity for the Maxwell algebra}},
  \href{https://doi.org/10.1007/JHEP10(2018)079}{\emph{JHEP} {\bfseries 10}
  (2018) 079} [\href{https://arxiv.org/abs/1805.08834}{{\ttfamily
  1805.08834}}].

\end{thebibliography}\endgroup
\bibliographystyle{JHEP.bst}
\end{document}